# Digital Twin of Electrical Tomography for Quantitative Multiphase Flow Imaging

Shengnan Wang,[1,2] Delin Hu,[1] Maomao Zhang,[3] Jiawang Qiu,[1] Wei Chen,[4] Francesco Giorgio-Serchi,[5] Lihui Peng,[4] Yi Li,[3] Yunjie Yang[1]*

[1] The SMART Group, Institute for Digital Communications, School of Engineering, The University of Edinburgh, Edinburgh, UK.
[2] College of Electrical, Energy and Power Engineering, Yangzhou University, Yangzhou, China.
[3] Tsinghua Shenzhen International Graduate School, Shenzhen, China.
[4] Department of Automation, Tsinghua University, Beijing, China.
[5] Institute for Integrated Micro and Nano Systems, School of Engineering, The University of Edinburgh, Edinburgh, UK.

**Abstract**

We report a digital twin (DT) framework of electrical tomography (ET) to address the challenge of real-time quantitative multiphase flow imaging based on non-invasive and non-radioactive technologies. Multiphase flow is ubiquitous in nature, industry, and research. Accurate flow imaging is the key to understanding this complex phenomenon. Existing non-radioactive multiphase flow imaging methods based on electrical tomography are limited to providing qualitative images. The proposed DT framework, building upon a synergistic integration of 3D field coupling simulation, model-based deep learning, and edge computing, allows ET to dynamically learn the flow features in the virtual space and implement the model in the physical system, thus providing unprecedented resolution and accuracy. The DT framework is demonstrated on gas-liquid two-phase flow and electrical capacitance tomography (ECT). It can be readily extended to various tomography modalities, scenarios, and scales in biomedical, energy, and aerospace applications as an effective alternative to radioactive solutions for precise flow visualization and characterization.

**Teaser**

Digital twin provides a unique pathway to customized, high-precision multiphase flow imaging with low-cost, non-radioactive electrical tomography.

## Introduction

Multiphase flow, as a transient and dynamic system with highly random, nonlinear, and hierarchical multi-scale natures, is prevalent in the natural environment, industrial processes and scientific research (*1-3*). Representative phenomena include blood flow in blood vessels (*4*), gas-liquid flow in post-combustion carbon capture processes (*5*), oil-gas flow in the energy industry (*6*), and micro-fluidic systems in biomedical research (*7*). A critical challenge in this field is quantitative visualization and characterization of the multiphase flow, which is vital to the fundamental study of flow mechanism, the prediction and control of flow behavior, process modeling, and the safe operation of industrial facilities (*8, 9*). Though some imaging techniques, e.g., X-ray tomography (*10-12*) and Magnetic Resonance Imaging (MRI) (*13*), can be used to provide quantitative flow images, their practicality is severely limited by their agility, scalability, cost, and radiological hazard. Electrical Tomography (ET), e.g., Electrical Capacitance Tomography (ECT) and Electrical Impedance Tomography (EIT), is considered a promising alternative for multiphase flow visualization and characterization (*14, 15*). It can provide an agile, non-invasive and non-radioactive way to unravel the time-varying distribution of the internal physical properties and high temporal resolution that facilitates the study of dynamic flow behavior at different scales and under extreme conditions (*14, 16*). Therefore, there has been tremendous interest in advancing ET for multiphase flow measurement over the last decades.

Despite advances in sensors, system design, and inverse problem theory, existing ET techniques are still inadequate for quantitative imaging of multiphase flows (*17, 18*). The underlying reason is the ill-posed and ill-conditioned nature of the ET inverse problem (*19*), leading to inevitable inversion errors. Another issue lies in the limited availability of ground truth data of fluid phases distribution for quantitative image evaluation (*20*). This is due to the highly complex nature of multiphase flows, which systematically prevents the time-history recording of accurate flow profiles.

Emerging deep learning and data-driven methods have the potential to resolve the nonlinear ET inverse problem (*21, 22*). Several learning-based imaging models, e.g., end-to-end learning (*23, 24*), model-based deep learning (*25, 26*), and unsupervised learning (*27, 28*), have been studied for high-resolution ET image reconstruction. The dataset plays a central role in these learning-based imaging approaches and determines the network's accuracy and generalization ability. Since the ground truth of multiphase flow profiles cannot be readily obtained in practice, the datasets of existing learning-based approaches are mainly constructed from static phantom data. Such static datasets are far from actual flow distributions and contain little information on dynamic flow behaviors, making learning-based models unfit to be transferred in realistic multiphase flow imaging scenarios.

We here propose a Digital Twin (DT) framework of ET to achieve quantitative imaging of multiphase flows by encapsulating dynamic 3D field coupling simulation, model-based deep learning, and edge computing. The DT framework is summarized in Fig. 1. The physical entity includes the testing section of a multiphase flow facility (Fig. 1A), the ET system (Fig. 1B), and the edge computer (Fig. 1C). A three-dimensional Fluid-Electrostatic field Coupling Model (3D-FECM) is developed as the digital representation of the physical multiphase flow imaging system. With 3D-FECM, the dynamic behavior of the real multiphase flows can be modelled, and instantaneous virtual ET measurements can be obtained simultaneously. By conducting dynamic coupling simulations, a virtual dataset consisting of tomographic data and corresponding flow profiles is generated (Fig. 1F). The framework also comprises a lightweight deep neural network (Fig. 1G), i.e., Deep Back Projection (DBP) (see Methods for details), trained based on the dataset and then implemented in the edge computer for quantitative multiphase flow imaging in the physical platform.

The DT framework provides an unprecedented way for ET to learn the dynamic flow features in the virtual space and enable quantitative multiphase flow imaging in the physical space. The DT framework is demonstrated on ECT and gas-liquid flow in this work but can be readily extended to other electrical tomography modalities, e.g., EIT or magnetic induction tomography, different multiphase flows, e.g., liquid-solid flow, and different scales. By adapting the coupling simulation model to specific cases, the DT framework also can be applied to other multiphase flow imaging techniques. This study provides a new paradigm for multiphase flow measurement, extends the limit of ET, and opens up a new avenue for developing artificial intelligence-based quantitative ET techniques.

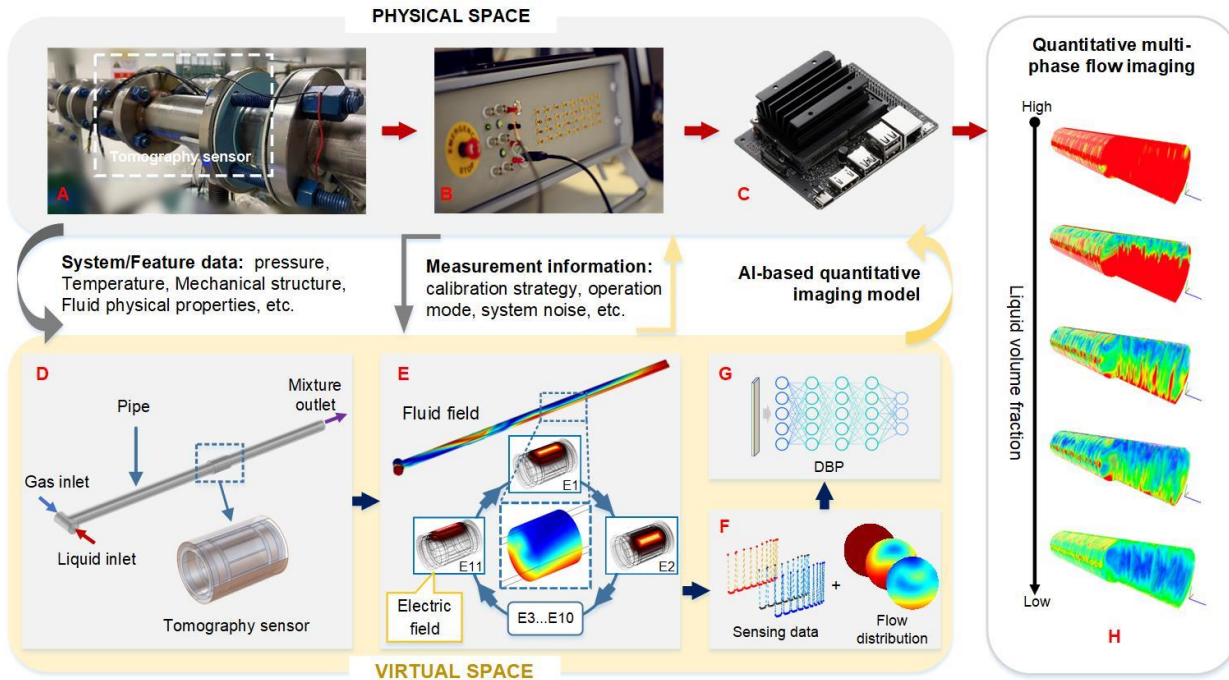

**Fig. 1. The digital twin framework of electrical tomography for quantitative multiphase flow imaging.** (**A**) The testing section of the multiphase flow facility. The system/feature data are captured via the sensors installed in the testing section. (**B**) The electrical tomography system. (**C**) The edge computer (i.e. NVIDIA Jetson Nano) implements the AI-based quantitative imaging model. (**D**) The 3D geometrical model of the testing section and the 12-electrode tomography sensor. (**E**) Schematic of the 3D fluid-electric field coupling simulation based on the model in (D). (**F**) Illustration of the generated virtual tomographic sensing data and corresponding dynamic flow distributions within the sensing region. (**G**) The model-based deep neural network (DBP) for quantitative flow imaging. DBP is trained based on the data set in (F) and implemented in (C). (**H**) The edge computer's output generates quantitative flow images and key parameters estimation based on the output of (B).

## Results

We first created a three-dimensional Fluid-Electrostatic field Coupling Model (3D-FECM) (see Methods for details of 3D-FECM and Fig. 2B) as the digital representation of the testing section of a pilot-scale multiphase flow facility (see Methods for multiphase flow facility details, Fig. S1 and Fig. 2A). In the multiphase flow facility, single-phase flows of gas (air) and liquid (white oil) are separately supplied and controlled to generate gas-liquid flows with different volumetric concentrations (see Fig. 2A). Similarly, in the virtual space, dynamic flows of gas and liquid are separately regulated to simulate various gas-liquid flows. Fig. 2C shows examples of typical sequential gas-liquid flows with 0.2 s intervals generated by the 3D-FECM. The pipe is initially filled with liquid. Gas-liquid flows are gradually formed in the horizontal pipe with the gas and liquid injection and then flow through the outlet. Additional representative gas-liquid flows generated from virtual space are presented in Movies S1 and S2.

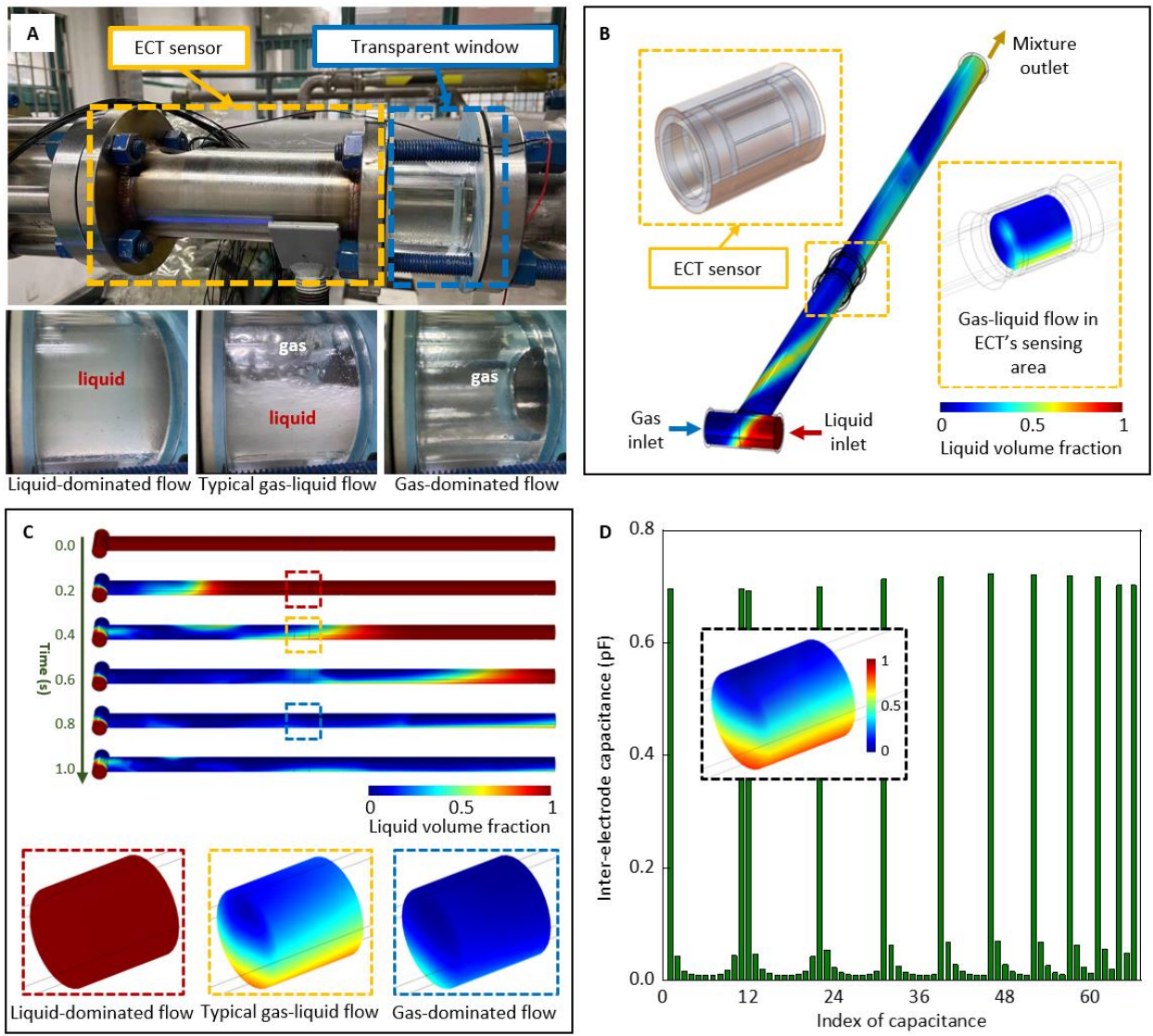

**Fig. 2. 3D coupling simulation of gas-liquid flows.** (**A**) Physical gas-liquid flows with different volumetric concentrations present in the testing section of the multiphase flow facility. (**B**) The schematic illustration of the 3D-FECM. Fluid field and electrostatic field are considered to simulate ECT measurements and dynamic gas-liquid flows. (**C**) Sequential gas-liquid two-phase flows generated by the 3D-FECM. (**D**) An example of the inter-electrode capacitance values obtained by the coupling simulation under a stratified flow.

By coupling the fluid and electrostatic fields, the specific electric potential distribution within the virtual ECT sensor is formed, and 66 independent inter-electrode capacitances can be obtained during the dynamic simulation process following the ECT measurement principle (*29*) (see Fig. 2D and Fig. S2). To imitate the Signal-Noise Ratio (SNR) of the real-world ECT system, which is around 60 dB (*30*), three levels of additive noise (SNR 60 dB, 50 dB, and 40 dB) are added to the virtual capacitances when reconstructing the cross-section liquid phase distributions in the sensor region. Referring to the actual working conditions of the multiphase flow testing facility, we conduct large-scale virtual experiments and synthesize a simulation dataset consisting of 12,362 samples of gas-liquid flow distributions and corresponding ECT measurements; see Methods for details of virtual multiphase flow data generation. Several examples of gas-liquid flow distributions and related images reconstructed using the conventional algorithm are shown in Figs. S3-4.

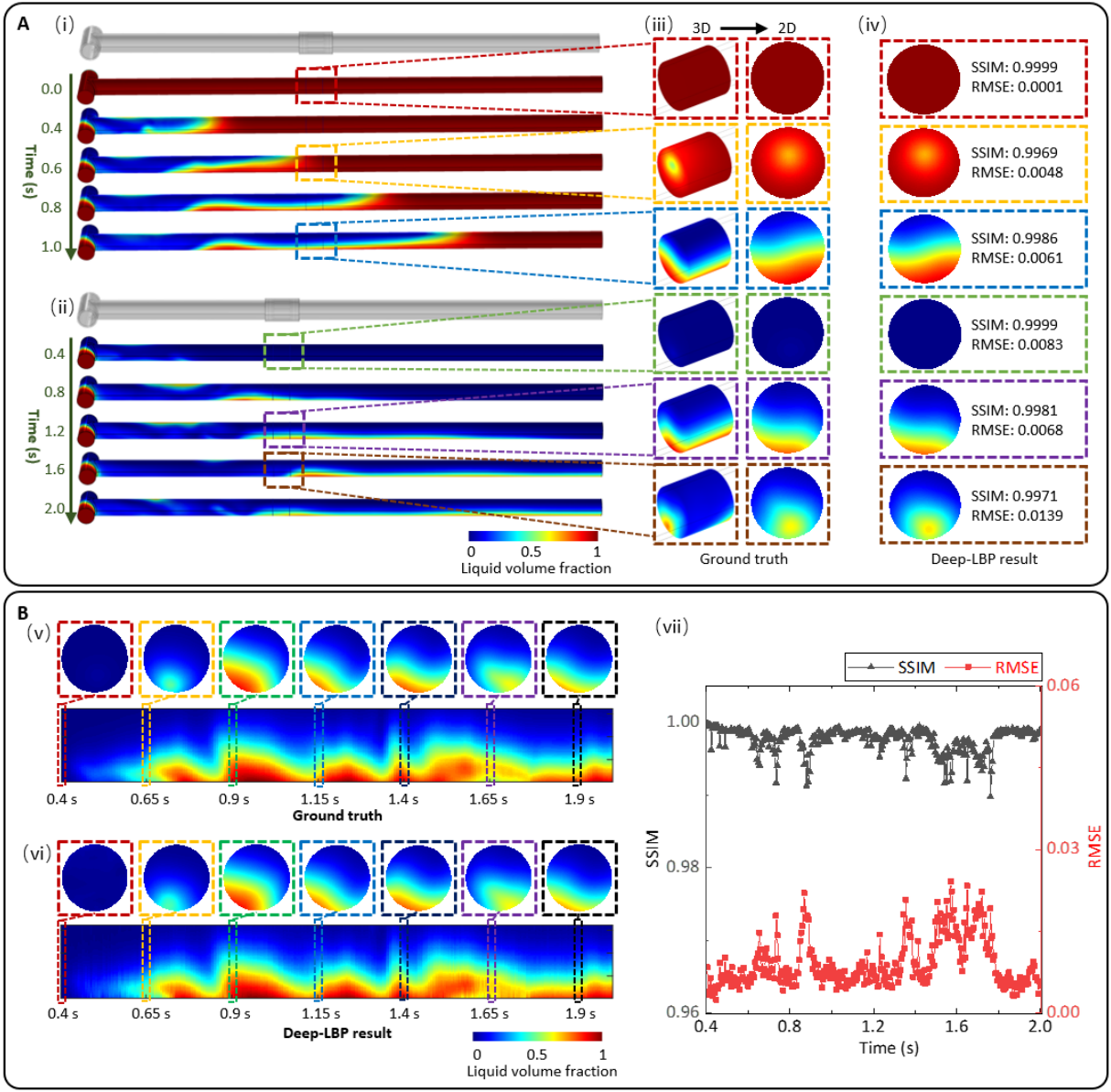

**Fig. 3. Quantitative imaging of virtual gas-liquid flows by DBP with 50 dB SNR.** (**A**) Two representative sets of sequential gas-liquid flows generated by 3D-FECM and corresponding image reconstruction results of DBP. For the sequential gas-liquid flows in (i), the pipe is initially filled with liquid. For the sequential gas-liquid flows in (ii), the pipe is initially filled with gas. 3D dynamic liquid phase distributions in the ECT sensing region can be converted to the 2D volume-averaged liquid phase distributions as the ground truth (see (iii)). The cross-section images reconstructed by DBP for the gas-liquid flows in (iii) are presented in (iv). (**B**) Quantitative imaging of virtual gas-liquid transient flow with high temporal resolution (200 frames per second). (v) The authentic liquid phase distributions for evaluating the performance of image reconstruction. (vi) The images reconstructed by DBP for the gas-liquid flows in (v). The cross-section images in (vi) are a set of representative images selected from the continuous reconstruction results. (vii) The SSIM and RMSE of the DBP results in (vi).

We develop a lightweight deep neural network (i.e., Deep Back Projection, DBP) and train the network using the simulation dataset (see Methods for details of DBP). We implement a series of virtual tests (using 50 dB noisy data) to verify the performance of the trained DBP for quantitative gas-liquid flow imaging. We calculate the 2D liquid phase distributions from the 3D liquid phase distributions of the ECT sensing region by averaging voxel-to-voxel along the axial direction of the sensor, and use them as the ground truth (see Fig. 3A). The Structural Similarity Index Measure (SSIM) (*31*) and the Root Mean Square Error (RMSE) (*32*) are adopted as the metrics to evaluate the reconstructed flow images quantitatively. Fig. 3A presents two representative sets of sequential gas-liquid flows generated by 3D-FECM and corresponding image reconstruction results using DBP when the pipe is initially filled with liquid and gas, respectively. The reconstructed cross-section images from both sets of sequential flows are close to the ground truth, with the SSIM higher than 0.997 and RMSE lower than 0.014. We implement virtual gas-liquid transient flow measurement with high temporal resolution (200 frames per second, 0.005 s intervals) to further examine the performance of DBP. Fig. 3B shows the image reconstruction results for the set of virtual sequential gas-liquid flows in Fig. 3A (ii).

Additionally, we also uniformly deploy eight virtual ECT sensors on the periphery of the pipeline to image the gas-liquid flows along the whole horizontal pipe section (see Fig. S6). The goal is to increase the diversity of flow patterns in the virtual dataset. The image reconstruction results in Fig. 3 and Fig. S9 show superior quality with the SSIM higher than 0.989 and RMSE lower than 0.024, indicating that the trained DBP can achieve accurate imaging of a wide variety of complex dynamic gas-liquid flows in the virtual space, which is not possible with the conventional ECT approach (see Figs. S7-9 for comparison).

We conduct gas-liquid dynamic flow imaging experiments on the pilot-scale multiphase flow facility as a case study to evaluate the performance of our DT framework (see Methods and Table S1 for detailed experimental setups/test matrix). Movies S3 presents gas-liquid flows captured by cameras under three experimental conditions, respectively. Some representative flows for the three experimental conditions are shown in Fig. 4A. The pipe is initially filled with liquid. With the gas and liquid injection, stratified gas-liquid flow is gradually presented in the horizontal section and flows through the ECT sensor. When the gas volume flow rate rises from 20.0 $m^3$/h to 100.0 $m^3$/h and the liquid volume flow rate drops from 5.0 $m^3$/h to 2.5 $m^3$/h, the liquid volumetric concentration of the gas-liquid flow in the testing section of the pipe decreases notably. Fig. 4B shows the continuous imaging results of the DT framework for each experimental condition. All the tomographic images present stratified flow, and the trend of fluid concentration for different conditions is in good agreement with that from the experimental flow observations. The experiment consists of three stable and two intermediate stages (see Fig. 4B(iv)). According to the tomographic data, the liquid volumetric concentration of the gas-liquid flow at each stage is $0.999 \pm 0.003$, $0.645 \pm 0.097$, and $0.221 \pm 0.113$ (mean ± standard deviation), respectively. The gas-liquid flows at intermediate stages contain more abundant dynamic features than the stable stages that are primarily stratified flows. Two sets of time-stacked tomographic images at intermediate stages are selected and presented in Figs. 4B (v) and (vi), respectively. The flow transitions show similarity compared with simulations, with the liquid volumetric concentration fluctuating within 0.988 to 0.730, and 0.735 to 0.306, respectively.

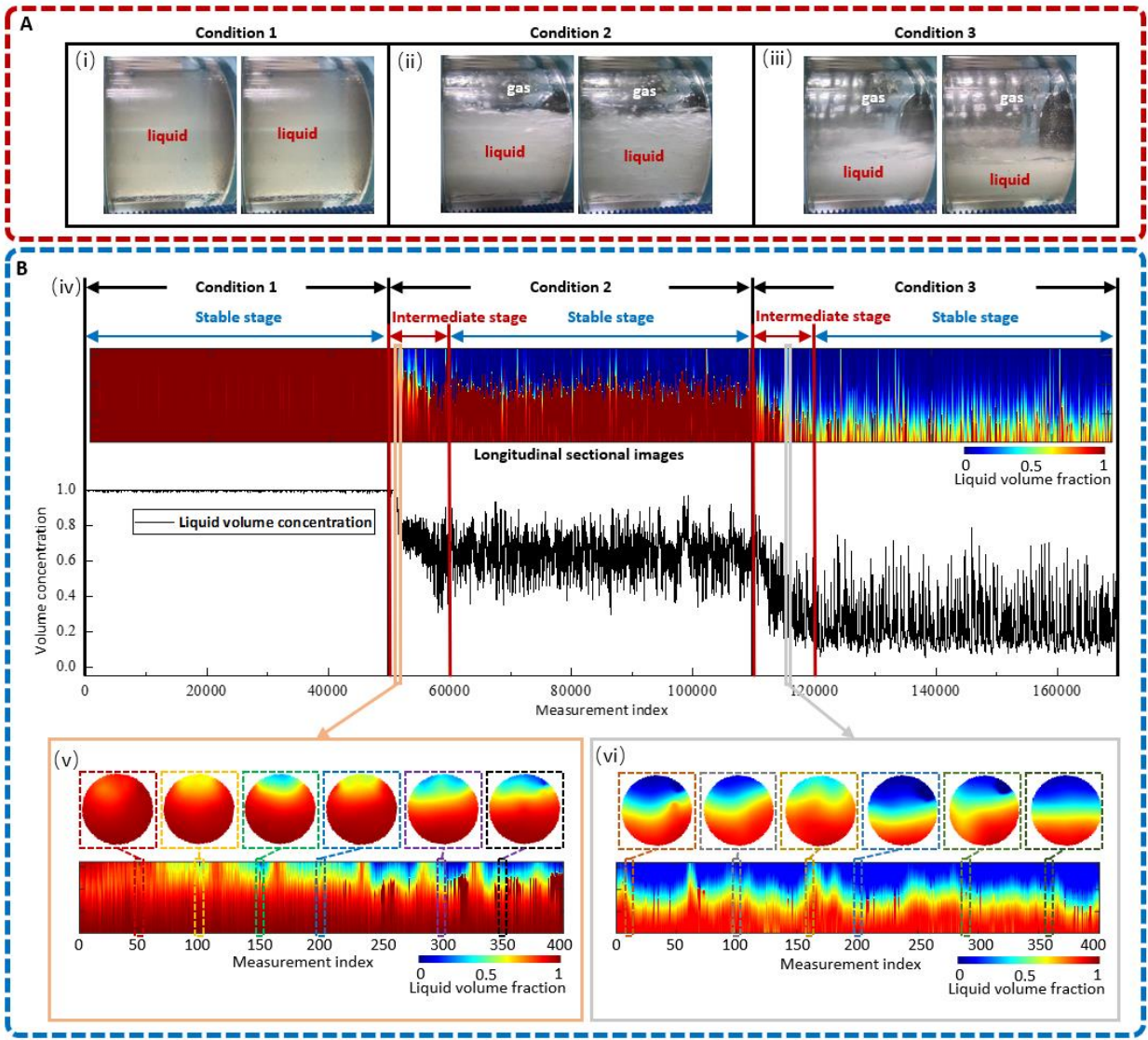

**Fig. 4. Tomographic images of gas-liquid flows in the pilot-scale multiphase flow facility under different experimental conditions.** In this experiment, three experimental conditions are selected for testing (see Methods for more details). (**A**) Flow profiles captured by camera. At the initial stage of the experiment, the pipe is filled with liquid (see (i)). The gas and liquid flow into the testing section with the volume flow rates of 20.0 m$^3$/h and 5.0 m$^3$/h, respectively. The gas-liquid two-phase flow with high liquid volumetric concentration is then formed (see (ii)). When the volume flow rate of the gas rises to 100.0 m$^3$/h and that of the liquid drops to 2.5 m$^3$/h, the gas-liquid flow with low liquid volumetric concentration is formed (see (iii)). (**B**) Tomographic images obtained from our DT framework. (iv) Tomographic images and liquid volumetric concentration variations of the gas-liquid flows in (A). Tomographic images in (v) and (vi) show two sets of representative images selected from the time-stacked cross-sectional reconstruction results in (iv).

We point out that the ground truth of dynamic multiphase flow profiles in practical flow facilities is mostly unavailable. As a result, quantitative evaluation of the reconstructed flow images has remained a long-standing yet unsolved challenge. We apply our DT framework to visualize and quantify static stratified gas-liquid flows in virtual and physical spaces. We also compare the real-world results with virtual-space results to provide an indicator of the

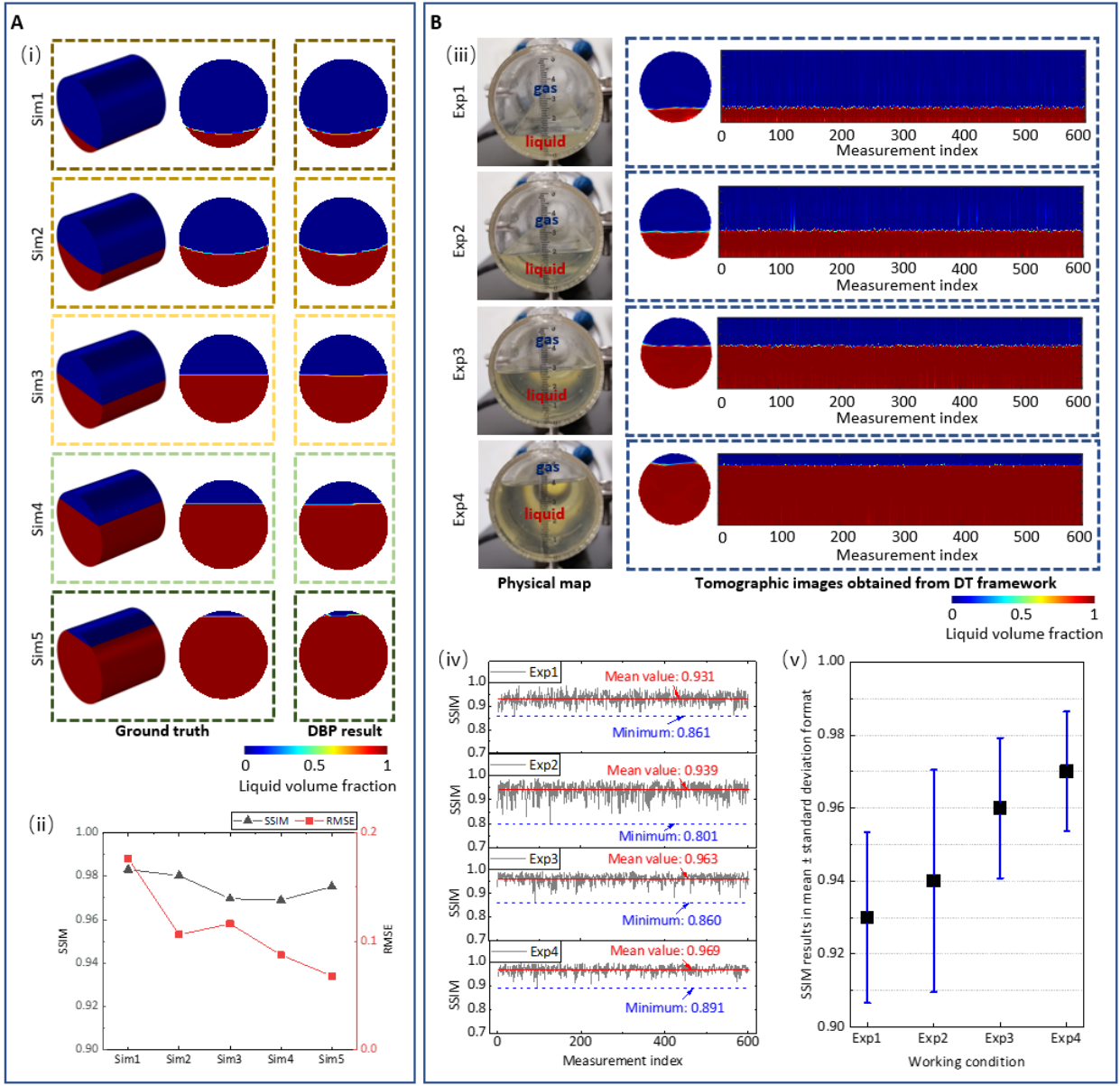

**Fig. 5. Evaluation of static stratified gas-liquid flow imaging in virtual and physical spaces.**
(**A**) Imaging static stratified gas-liquid flows in virtual space by DBP with 50 dB data. (i) Virtual static gas-liquid flows and corresponding image reconstruction results of DBP. (ii) The SSIM and RMSE of DBP results in (i). (**B**) Imaging static stratified gas-liquid flows in physical space using DBP. (iii) Real-world static gas-liquid flows and corresponding image reconstruction results of DBP. (iv) The SSIM of the DBP results in (iii). (v) The standard uncertainty of our DT framework for imaging the real-world static gas-liquid flows in (iii).

feasibility and performance of the DT framework in the physical world from a quantitative perspective. Figs. 5A and B show the imaging results from the virtual and physical spaces, respectively. We observe that the tomographic images of the virtual static stratified flows and real-world flows are very close to actual distributions. The SSIMs of the images for the virtual static flows are higher than 0.969, and those for the real-world static flows are larger than 0.801, indicating that the DT framework can accurately visualize the gas-liquid flows both in virtual and physical spaces. From the continuous imaging results of gas-liquid flows in the physical facility (see Fig. 5B), we also see that the SSIM result for each working

condition is 0.931 ± 0.023, 0.939 ± 0.030, 0.963 ± 0.019, 0.969 ± 0.016 (mean ± standard deviation), respectively. The relative standard uncertainty of the imaging results is better than 3.19%, indicating superior measurement stability and high repeatability of the DT framework.

**Discussion**

In this study, we first introduce the DT concept to multiphase flow imaging systems. We propose a DT framework for ET and demonstrate that it can effectively learn the multiphase flow features in virtual space and provide unprecedented resolution and accuracy of multiphase flow imaging in physical space. Despite the superiority of our DT framework, several limitations exist due to current technical bottlenecks and theoretical defects.

In virtual space, we leverage 3D-FECM to build the digital representation of the physical multiphase flow imaging system. Coupling the fluid and electrical fields allows simultaneous simulation of dynamic multiphase flows and imaging sensors. However, it is noteworthy that the fundamentals and mathematical treatment of multiphase flow modelling are still largely undeveloped (*33*). Real-world multiphase flow is a highly complex system whose flow types and flow regimes within each flow type can be significantly affected by various conveying conditions, such as small vessel deformations, operating errors, and control valve disturbances. Achieving a perfect match between the simulation and reality remains a critical challenge. To mitigate this issue, instead of focusing on the accuracy of the 3D-FECM model to replicate physical multiphase flows, our attention shifts to generating a variety of gas-liquid flows carrying abundant dynamic features to cover a wide range of complex flows in the flow facility. To promote computational efficiency, we simplified the 3D-FECM to simulate the testing section rather than the whole facility. We also performed additional multiphase flow simulations under the microgravity environment to achieve a broader coverage of flow types and flow regimes in the virtual dataset. Another limitation of the virtual model is that, although only the testing section is modeled, it still takes tremendous time to generate 3D virtual flow and tomographic measurements for each working condition. This is detrimental to the real-time performance of the DT framework. Model optimization to significantly reduce computational cost and leveraging more powerful computing hardware could be a potential solution.

In physical space, we applied our DT framework to visualize dynamic gas-liquid flows in the laboratory-scale multiphase flow facility. However, it was extremely challenging to obtain the ground truth of dynamic gas-liquid flow profiles that could be used for quantitative performance evaluation. Alternatively, we quantitatively evaluated the imaging results of static stratified flows and compared the dynamic imaging results with those captured by high-speed cameras. Static imaging results reveal that the DT framework can visualize real-world gas-liquid flows with high accuracy and excellent repeatability. Future improvements will be to benchmark the experimental performance by incorporating other advanced imaging techniques, e.g., X-ray tomography and MRI, and to compare DT results with these high-precision imaging techniques. Nevertheless, this will involve multi-sensors integration and multi-signal fusion, and the implementation will be complicated and challenging.

It is also noteworthy that the focus of this work is the overall DT framework rather than the learning-based algorithm for ECT image reconstruction. We demonstrate that our DT framework can achieve superior performance even using simple neural architectures like DBP. We also point out that employing dedicated and more complicated neural architectures can potentially lead to a better performance at the cost of a larger training dataset and a more complex training strategy.

In summary, the proposed DT framework for ET utilizes 3D field-coupling simulation, model-based deep learning, and edge computing to enable precise flow profiles imaging with low-cost, non-radiative, and non-invasive tomography techniques. We demonstrated substantial improvements of DT-powered ET over conventional ET both virtually and in a pilot-scale multiphase flow facility under various gas-liquid flow conditions. Our DT framework can be trained efficiently and flexibly in the virtual space and be readily implemented in the physical space to provide quantitative and stable imaging of gas-liquid flows, representing a step change compared to the state of the art. The framework is in principle generalizable to various imaging techniques, and emerging real-time simulation/data sketching techniques could realistically propel our DT framework towards widespread multiphase flow imaging applications.

## Materials and Methods

### Multiphase Flow Facility and Experiment Design

The testing section of a pilot-scale multiphase flow facility (see Fig. S1) at the Multiphase Flow Engineering Laboratory of the Tsinghua International Graduate School is adopted as a case study. The facility consists of a multiphase flow separator, a gas storage tank, gas and liquid single-phase flow sections, the mixing section, and the control system. In this study, the gas-liquid two-phase flow is considered. The working gas and liquid are air (permittivity 1.0, density 1.3 kg/m$^3$) and white oil (permittivity 2.18, density 879 kg/m$^3$), respectively. The oil is separately supplied and pumped into the flow pipe, then blended with the gas in operation. The mixture is transported through the gas-liquid flow testing section and returned to the separator for circulating utilization. A 12-electrode ECT sensor with a transparent window for visual observation is installed in the horizontal mixing section (see Fig. S1C). The testing section has an internal diameter of 50 mm, in which gas-liquid two-phase flows with different volumetric concentrations can be formed by manipulating the air and white oil volume flow rates. During the experiment, the working pressure in the testing section is set to 0.6 MPa and the experimental temperature is about 33 ℃. The experimental conditions are listed in Table S1, where the volume flow rate of white oil varies from 5.0 to 2.5 m$^3$/h, and the volume flow rate of air ranges from 20.0 to 100.0 m$^3$/h.

### 3D Field Coupling Simulation

We created a three-dimensional Fluid-Electrostatic field Coupling Model (3D-FECM) to duplicate the testing section of the flow facility. 3D field coupling simulation was performed using the commercial software COMSOL Multiphysics and Matlab. The model contains the fluid field interface to generate the gas-liquid flow data and an electrostatic field interface to simulate the 12-electrode ECT sensor. Fig. S2 shows the flowchart of the 3D field coupling simulation. We can simultaneously obtain the dynamic permittivity distribution under various flow conditions and corresponding capacitance measurements from the virtual ECT sensor by coupling the fluid field and the electrostatic field.

For the fluid field interface, we employ the laminar two-phase flow, level set method (*34*) to track the moving interface between the gas and liquid phases. The gas and liquid phases are air and white oil, respectively. We impose the velocity inlet and pressure outlet boundary conditions to avoid convergence difficulties. We select the suppress backflow to prevent fluid from entering the domain through the outlet boundary. The gas-liquid two-phase flow is set as incompressible flow. In line with the experimental setup, the temperature of the simulation environment is set as 33 ℃, the dynamic viscosity of the gas and liquid is set as 1.81E-5 Pa.s and 0.02 Pa.s, respectively, and the density of the gas and liquid is set as 1.3 kg/m$^3$ and 879 kg/m$^3$, respectively.

We apply the Poisson equation (*35*) to determine the electric potential distribution for the electrostatic field interface. We then use the Wiener Upper Bound formula (*36*) to evaluate the equivalent permittivity of the gas-liquid mixture. The relative permittivity of the pipe, gas, and liquid is set as 2.6, 1.0, and 2.18, respectively. All the 66 non-redundant inter-electrode capacitances are collected for image reconstruction.

**Virtual Multiphase Flow Dataset Generation**
To replicate the real scenarios of the gas-liquid flow testing facility, we initially apply the incompressible Navier-Stokes equations (*37*) with gravity to simulate gas-liquid two-phase flows in the horizontal section. The dynamic gas and liquid single-phase flows are separately supplied and controlled. Gas-liquid two-phase flow data with different volumetric concentrations are generated by regulating the inlet velocities perpendicular to the entrance surfaces. The inlet liquid velocity varies from 0.071 to 0.708 m/s, and the inlet gas velocity varies from 0.236 to 2.362 m/s. To cover a wide range of volumetric concentrations, two initial conditions are set. One is that the pipe is filled with liquid at the initial stage, and the other is that the pipe is filled with gas at the initial stage. We conduct additional virtual dynamic experiments to generate gas-liquid two-phase flows in the microgravity environment to obtain more abundant flow regimes for the machine learning model training. Table S2 gives the virtual dynamic experimental matrix. A working condition for the virtual static experiment is also added to generate static stratified gas-liquid flow with the liquid volumetric concentration ranging from 0 to 1. We ran 74 working conditions and collected 12,362 virtual samples containing ECT measurements and phase distributions.

**DBP for Quantitative Flow Imaging**
We found that the conventional Linear Back Projection (LBP) algorithm (*29*) is the most effective in reconstructing dynamic flow profiles than iterative algorithms. We therefore further refine the LBP results with machine learning. DBP is a model-based deep learning algorithm designed to quantitatively image the flow profiles (see Fig. S5 for the network structure). Here, we consider the commonly used ECT model (*29*), i.e.

$$\lambda = Sg \qquad (1)$$

where $\lambda$ denotes the normalized capacitance measurement vector; $S$ is the Jacobian matrix and $g$ represents the normalized permittivity distribution within the Region of Interest (ROI).

Reconstruction of the flow distribution with DBP involves two steps. First, the normalized measurement vector $\lambda$ is mapped into a coarse flow distribution $g_{LBP}$ based on the LBP algorithm (*29*). Then, a modified UNet (*38*) is applied to refine the LBP result and produce a more accurate image $g_U$.

DBP is implemented in Pytorch. The Adam optimizer (*39*) is used to update network parameters. The hyperparameters of Adam are set as: $\beta_1$=0.9, $\beta_2$=0.999, $\epsilon$=10$^{-9}$, weight decay=0. The initial learning rate is 0.001, which decays every 2 epochs with a factor of 1.111. The simulation data is divided into three groups, i.e., the training, validation, and testing sets. The training set includes 10,505 samples (62 different flow conditions at normal time resolution in the dynamic simulation and 89 different flow conditions in the static simulation). The validation set includes 1,680 samples (10 different flow conditions at normal time resolution in the dynamic simulation). The testing set includes 1,380 samples (one flow condition at normal time resolution, three flow conditions at high time resolution in the dynamic simulation, and nine different flow conditions in the static simulation). All data is augmented by 3 noise levels (i.e., 40dB, 50dB and 60dB). We select mean square error as the loss function. The batch size and the number of epochs are set to 25 and 80, respectively. The whole training takes about 3 hours on three Nvidia Quadro P5000 GPUs. The network with the least validation loss is selected as our final model, which can achieve $0.996 \pm 0.012$ (mean ± standard deviation) for SSIM, $0.005 \pm 0.008$ (mean ± standard deviation) for RMSE, and $40.218 \pm 1.701$ (mean ± standard deviation) for Peak Signal to Noise Ratio (PSNR) on the whole testing set.

**AI-powered Tomography System**

The trained DBP is implemented in the AI-powered electrical tomography system in the physical space. We use ECT to demonstrate the construction of the AI-power tomography system. However, the architecture could be easily extended to other electrical tomography modalities. The AI-powered ECT system is composed of a 32-channel ECT hardware (*30, 40*), an edge AI computer (NVIDIA Jetson Nano), and a Visual Tomography (VT) software integrating the trained DBP model for real-time quantitative flow profile reconstruction and key parameter prediction (see Fig.S10 for the system architecture). The ECT hardware is interfaced with the Jetson Nano through a USB2.0 port. The VT software developed via Python is implemented on Jetson Nano for ECT measurement control, data collection, image reconstruction, and visualization. It also provides an interface to update the trained DBP model dynamically and remotely.

## Acknowledgments

**Funding:** S.W., D.H. and Y.Y. is supported by:
European Union's Horizon 2020 Research and Innovation Programme under the Marie Sklodowska-Curie actions COFUND Transnational Research and Innovation Network at Edinburgh Grant 801215
LeEngStar Technology Co., Ltd
Data Driven Innovation Chancellor's Fellowship
National Natural Science Foundation of China Grant 51906209

**Author contributions:** Y.Y. conceived the idea and supervised the entire project. S.W. developed the 3D field coupling simulation framework, generated the virtual sensing data, carried out data analysis, and led the the writing of the manuscript. D.H. and J.Q. developed the DBP network and conducted data analysis. M.Z. and Y.L. managed experiments and data collection. W.C. conducted static experiments. F.G.S. supervised the fluid simulation. L.P. supervised the electrical tomography parts and measurement system design. All the authors participated in discussions of the results and contributed to the writing of the manuscript.

**Competing interests:** Authors declare that they have no competing interests.

**Data and materials availability:** All data and code will be available from the corresponding author upon reasonable request.